\documentclass[12pt,a4paper]{article}
\input{tcilatex}
\begin{document}

\begin{center}
\vspace{1pt}\textbf{Determination of Low-Energy Parameters of
Neutron--Proton Scattering}

\textbf{on the Basis of Modern Experimental Data from Partial-Wave Analyses}
\end{center}

\textbf{\vspace{1pt}}

\begin{center}
\textbf{V. A. Babenko\footnote{%
E-mail: pet@online.com.ua} and N. M. Petrov}\\[0pt]
\textit{Bogolyubov Institute for Theoretical Physics, National Academy of
Sciences of Ukraine,}

\textit{Metrologicheskaya ul. 14b, 03143 Kiev, Ukraine\vspace{1pt}\vspace{1pt%
}}
\end{center}

\vspace{1pt}

\noindent The triplet and singlet low-energy parameters in the
effective-range expansion for neutron--proton scattering are determined by
using the latest experimental data on respective phase shifts from the SAID
nucleon--nucleon database. The results differ markedly from the analogous
parameters obtained on the basis of the phase shifts of the Nijmegen group
and contradict the parameter values that are presently used as experimental
ones. The values found with the aid of the phase shifts from the SAID
nucleon--nucleon database for the total cross section for the scattering of
zero-energy neutrons by protons, $\sigma _{0}=20.426\,$b, and the
neutron--proton coherent scattering length, $f=-3.755\,$fm, agree perfectly
with the experimental cross-section values obtained by Houk, $\sigma
_{0}=20.436\pm 0.023\,$b, and experimental scattering-length values obtained
by Houk and Wilson, $f=-3.756\pm 0.009\,$fm, but they contradict
cross-section values of $\sigma _{0}=20.491\pm 0.014\,$b according to Dilg
and coherent-scattering-length values of $f=-3.7409\pm 0.0011\,$fm according
to Koester and Nistler.

\vspace{12pt}\vspace{1pt}

\vspace{1pt}PACS: 13.75.Cs, 21.30.-x, 25.40.Dn

\textbf{DOI}: 10.1134/S1063778807040072

\textbf{1.} Along with the deuteron parameters, the low-energy parameters in
the effective-range expansion for neutron--proton scattering,
\begin{equation}
k\cot \delta =-\frac{1}{a}+\frac{1}{2}%
rk^{2}+v_{2}k^{4}+v_{3}k^{6}+v_{4}k^{8}+\ldots \,,  \label{1}
\end{equation}
\vspace{1pt}are fundamental quantities that play a key role in studying
strong nucleon--nucleon interaction. These parameters are of great
importance for constructing various realistic nuclear-force models, which,
in turn, form a basis for studying the structure of nuclei and various
nuclear processes. For this reason, it is highly desirable to determine
reliably and accurately the parameters in the effective-range expansion,
including the scattering length $a$, the effective range $r$, the shape
parameter $v_{2}$, and higher order parameters $v_{n}$.

Although low-energy parameters for neutron--proton scattering have been
determined and studied since the early 1950s, even the experimental values
of such parameters as the scattering length $a$ and the effective range $r$
are ambiguous to date. As for the shape parameter $v_{2}$, even its sign is
unknown at the present time. The theoretical value of this parameter depends
greatly on the nuclear-force model used: as we go over from one model to
another, the parameter $v_{2}$ in the triplet state changes within a broad
interval, from $-0.95$ \lbrack 1, 2\rbrack\ to $1.371\,$fm$^{3}$ \lbrack
3\rbrack , whence it follows that the shape parameter is a very subtle and
sensitive feature of nucleon--nucleon interaction.

We would like to note that not only does the shape parameter $v_{2}$ depend
on the form of interaction, but it is also strongly dependent on the
scattering length $a$ and the effective range $r$. In particular, a change
of only a few tenths of a percent in the scattering length $a$ may lead to a
severalfold change in the shape parameter $v_{2}$ \lbrack 4\rbrack . The
shape parameters $v_{n}$ of order higher than that of $v_{2}$ have been
still more poorly determined and are more sensitive to details of
nucleon--nucleon interaction. The aforesaid highlights once again the
importance of reliably determining the scattering length $a$ and the
effective range $r$, the more so as these are quantities that are most
frequently used as inputs in constructing various models of nucleon--nucleon
interaction.

\textbf{2.} It is well known \lbrack 5\rbrack\ that the neutron--proton
system may occur either in the triplet (the total spin is $S=1$) or the
singlet (the total spin is $S=0$) spin state. In determining the scattering
lengths $a$ and the effective ranges $r$ in the triplet ($t$) and singlet ($%
s $) spin states, one employs the experimental dependence of the total
(spin-averaged) cross section for the scattering of slow neutrons by free
protons and data characterizing the scattering of zero-energy neutrons by
para-hydrogen. In order to determine the triplet and singlet scattering
lengths ($a_{t}$ and $a_{s}$, respectively), use is usually made of
equations that relate these quantities to the total cross section for the
scattering of zero-energy neutrons by protons,
\begin{equation}
\sigma _{0}=\pi \left( 3a_{t}^{2}+a_{s}^{2}\right) \,,  \label{2}
\end{equation}
and to the coherent scattering length,
\begin{equation}
f=\frac{1}{2}\left( 3a_{t}+a_{s}\right) \,.  \label{3}
\end{equation}

\vspace{1pt}

In this case, the cross section $\sigma _{0}$ is determined from the results
of experiments that study slow-neutron scattering on protons bound in
various molecules (H$_{2}$, H$_{2}$O, C$_{6}$H$_{6}$, CH$_{3}$OH),
corrections associated with neutron capture by a proton and with effects of
proton binding in molecules being subsequently eliminated. The elimination
of binding-effect corrections is a nontrivial many-body problem, since, in
addition to proton and neutron motion, it is necessary to take into account
the motion of the molecular residue. A number of significant simplifications
and approximations are made in solving this problem \lbrack 6\rbrack . A
compendium of experimental results from \lbrack 7--13\rbrack\ on the total
cross section for the scattering of zero-energy neutrons by free protons, $%
\sigma _{0}$, is given in Table 1.

Two values of the total cross section $\sigma _{0}$ are recommended at the
present time. These are the value obtained by Houk (1971) \lbrack 12\rbrack
,
\begin{equation}
\sigma _{0}=20.436(23)\,\text{b,}  \label{4}
\end{equation}
and the value obtained by Dilg (1975) \lbrack 13\rbrack ,
\begin{equation}
\sigma _{0}=20.491(14)\,\text{b.}  \label{5}
\end{equation}
Since these two values of $\sigma _{0}$ are inconsistent, their
weighted-mean value
\begin{equation}
\sigma _{0}=20.476(12)\,\text{b}  \label{6}
\end{equation}
can also be used in determining the scattering lengths.

\vspace{1pt}

It should be noted that the total cross section $\sigma _{0}$ has not been
measured since 1975.

\vspace{1pt}\vspace{1pt}

The coherent scattering length $f$, which is determined by relation (3), is
found either from experiments where slow neutrons are scattered by pure
para-hydrogen \lbrack 8, 14, 15\rbrack\ or by crystals \lbrack 16\rbrack\ or
--- and this is a more precise method --- from experiments where neutrons
are reflected by a liquid mirror and where use is made of a number of pure
hydrocarbons \lbrack 9, 10, 17--22\rbrack . Also, a method for determining
the coherent scattering length by means of neutron interferometry from
experiments to study neutron scattering on molecular hydrogen was proposed
in \lbrack 23\rbrack . The values found by various authors for the
neutron--proton coherent scattering length $f$ are quoted in Table 2, whence
it can be seen that the value of this quantity is even more ambiguous than
the value of $\sigma _{0}$.

In determining the scattering lengths in the triplet and the singlet state ($%
a_{t}$ and $a_{s}$, respectively), one employs most frequently, at the
present time, the coherent-length value obtained by Koester and Nistler
\lbrack 22\rbrack ,
\begin{equation}
f=-3.7409(11)\,\text{fm,}  \label{7}
\end{equation}
and the coherent-length value presented in the compilation of Dumbrajs et
al. \lbrack 24\rbrack ,
\begin{equation}
f=-3.738(1)\,\text{fm.}  \label{8}
\end{equation}
Recent experiments aimed at determining the neutron--proton coherent
scattering length by means of neutron interferometry \lbrack 23\rbrack ,
which were mentioned above, yielded the value
\begin{equation}
f=-3.7384(20)\,\text{fm.}  \label{9}
\end{equation}
Within the experimental errors, the value in (9) agrees with the result of
Koester and Nistler in (7) and with the value in (8), which was used by
Dumbrajs et al. \lbrack 24\rbrack .

Table 3 presents values obtained in a number of previous studies \lbrack 9,
10, 13, 18, 21, 22, 24--28\rbrack\ for the scattering lengths and effective
ranges in the triplet and singlet spin states. All of them have been used as
experimental values. The values of the triplet ($a_{t}$) and singlet ($a_{s}$%
) scattering lengths from Table 3 were obtained on the basis of formulas (2)
and (3) by using various values for the total cross section $\sigma _{0}$
and the neutron--proton coherent scattering length $f$.

The values of the triplet effective range $r_{t}$ in Table 3 were determined
primarily in an approximation that does not depend on the form of
interaction; that is, 
\begin{equation}
r_{t}\equiv \rho \left( -\varepsilon _{d},0\right) =2R\left( 1-\frac{R}{a_{t}%
}\right) \,,  \label{10}
\end{equation}
where\ $\rho \left( -\varepsilon _{d},0\right) $ is the mixed effective
radius of the deuteron; 
\begin{equation}
R=1/\alpha  \label{11}
\end{equation}
is a parameter that characterizes the spatial dimensions of the deuteron;
and $\alpha $ is the deuteron wave number, which is related to the deuteron
binding energy $\varepsilon _{d}$ by the equation 
\begin{equation}
\varepsilon _{d}=\hbar ^{2}\alpha ^{2}/m_{N}\,.  \label{12}
\end{equation}

In a number of studies \lbrack 24, 26\rbrack , the triplet effective range
was determined in accordance with the formula 
\begin{equation}
r_{t}=\rho \left( -\varepsilon _{d},0\right) +\delta r_{t}\,,  \label{13}
\end{equation}
where the correction $\delta r_{t}$ is a model-dependent quantity. According
to the estimates obtained by Noyes on the basis of the dispersion relations
\lbrack 26\rbrack , the correction $\delta r_{t}$ arising owing to one-pion
exchange is 
\begin{equation}
\delta r_{t}=-0.013\,\text{fm}\,.  \label{14}
\end{equation}
According to other estimates \lbrack 24\rbrack , this correction is 
\begin{equation}
\delta r_{t}\simeq -0.001\,\text{fm}\,,  \label{15}
\end{equation}
which is an order of magnitude smaller in absolute value than the estimate
in (14). In the latter case, the effective range $r_{t}$ is therefore nearly
coincident with the mixed effective radius $\rho \left( -\varepsilon
_{d},0\right) $.

\vspace{1pt}

The singlet effective range $r_{s}$ is usually determined on the basis of an
analysis of the total cross section for neutron--proton scattering, $\sigma
\left( E\right) $, in the low-energy region at fixed values of the
parameters $a_{t}$, $a_{s}$, and $r_{t}$. The values found in this way for
the singlet effective range $r_{s}$ appear to be even more ambiguous than
the values of the triplet effective range. As can be seen from Table 3, the
scattering-length and effective-range values used as experimental ones
change within rather broad ranges. The scatter of these values is due first
of all to the fact that different experimental values of the cross section
for the scattering of zero-energy neutrons by free protons, $\sigma _{0}$,
and of the neutron--proton coherent scattering length $f$ are used to
determine these quantities. The ambiguity in determining the singlet
effective range $r_{s}$ is also associated with an insufficient accuracy of
the experimental total cross sections for neutron--proton scattering at
energies below $5\,$MeV. The values found by different authors for the
singlet effective range $r_{s}$ change within a broad range, from $2.42$
\lbrack 9\rbrack\ to $2.81\,$fm \lbrack 13\rbrack .

Thus, the accuracy of the experiments performed in the 1950s--1970s is
insufficient for unambiguously determining the low-energy parameters of
neutron--proton scattering. At the same time, these parameters play an
important role in the theory of few-nucleon systems, which is based on
nucleon--nucleon interaction. As was shown in \lbrack 29, 30\rbrack , the
binding energies of the $^{3}$H and $^{4}$He nuclei depend greatly on the
singlet effective range $r_{s}$, increasing as $r_{s}$ becomes smaller. By
way of example, we indicate that, as $r_{s}$ decreases by $0.1\,$fm, the
binding energies of the $^{3}$H and $^{4}$He nuclei increase by $0.3$ and $%
1.5\,$MeV, respectively. We note that the decrease of $0.01\,$fm in the
triplet scattering length $a_{t}$ also leads to the increase of $0.025\,$MeV
in the triton binding energy \lbrack 31, 32\rbrack . At the same time, it is
well known that, in calculations with realistic nucleon--nucleon potentials,
the binding energies of few-nucleon systems prove to be underestimated. In
such calculations, the $^{3}$H binding energy is as a rule underestimated by 
$1\,$MeV. A reliable and precise determination of the low-energy parameters
of neutron--proton scattering and their use in calculating the binding
energies of systems that contain three or more nucleons may contribute to
solving the problem of underestimating the binding energies of few-nucleon
systems without introducing three-particle forces, quark degrees of freedom,
and other concepts that would require revising basic points in the
traditional theory of nuclear forces, which relies on pair nucleon--nucleon
interaction.

To conclude this section, we present, for low-energy parameters, values that
are currently used as experimental ones. Most frequently, the present-day
literature quotes two sets of low-energy parameters. These are the set from
\lbrack 24\rbrack , 
\begin{equation}
\begin{array}{c}
a_{t}=5.424(4)\,\text{fm, }r_{t}=1.759(5)\,\text{fm;} \\ 
a_{s}=-23.748(10)\,\text{fm, }r_{s}=2.75(5)\,\text{fm,}
\end{array}
\label{16}
\end{equation}
which is matched with the experimental value (5) of the total cross section
at zero energy due to Dilg \lbrack 13\rbrack\ and with the value in (8) for
the neutron--proton coherent scattering length from \lbrack 24\rbrack , and
the set from \lbrack 28\rbrack , 
\begin{equation}
\begin{array}{c}
a_{t}=5.419(7)\,\text{fm, }r_{t}=1.753(8)\,\text{fm;} \\ 
a_{s}=-23.740(20)\,\text{fm, }r_{s}=2.77(5)\,\text{fm,}
\end{array}
\label{17}
\end{equation}
which corresponds to the weighted-mean value (6) of the cross sections
presented by Houk \lbrack 12\rbrack\ and Dilg \lbrack 13\rbrack\ and to the
value in (7) for the coherent length due to Koester and Nistler \lbrack
22\rbrack . It should be noted that the experiments performed in the
1950--1970s were the main source of information used to deduce the values in
(16) and (17) for the low-energy parameters of neutron--proton scattering.

\textbf{3.} In recent years, the accuracy of experimental data on
nucleon--nucleon scattering has been improved considerably; moreover,
methods of their partial-wave analysis, which make it possible to describe
the results of scattering experiments in terms of phase shifts, have also
been refined \lbrack 33, 34\rbrack . Owing to this, the triplet and singlet
low-energy parameters of neutron--proton scattering can be determined
independently of one another by using the $^{3}S_{1}$- and $^{1}S_{0}$-state
phase shifts \lbrack 4, 35\rbrack . The results of the partial-wave analysis
performed by the GWU group \lbrack 33\rbrack\ (data from the well-known SAID
nucleon--nucleon database) and by the Nijmegen group \lbrack 34\rbrack\ are
presently the most precise and most widely used data on the phase shifts for
nucleon--nucleon scattering. The most popular modern realistic
nucleon--nucleon potentials constructed within the last decade, which
include the Nijm-I, Nijm-II, Reid93 \lbrack 36\rbrack , Argonne V18 \lbrack
37\rbrack , CD-Bonn \lbrack 28, 38\rbrack , and Moscow \lbrack 39\rbrack\
potentials, are based on fits to data of the Nijmegen group \lbrack
34\rbrack . However, it should be noted that the partial-wave analysis of
the Nijmegen group is a result of processing and averaging experimental data
on nucleon--nucleon scattering over a period from 1955 to 1992, but this
analysis provides an insufficiently accurate description of modern
experimental data on nucleon--nucleon scattering. Despite the proximity of
the phase shifts for neutron--proton scattering that were obtained by the
GWU and Nijmegen groups, the corresponding values of the low-energy
parameters in the effective-range expansion are markedly different \lbrack
4\rbrack , this difference being not only quantitative but also qualitative.

Using the approximation of the effective-range function $k\cot \delta $ at
low energies by polynomials and Pad\'{e} approximants within the least
squares method, we calculated the triplet and singlet low-energy parameters
of neutron--proton scattering for the experimental data on the GWU \lbrack
33\rbrack\ and Nijmegen \lbrack 34\rbrack\ phase shifts. The results
obtained for the low-energy parameters in the present study by employing the
data from the partial-wave analysis of the GWU group, 
\begin{equation}
\begin{array}{c}
a_{t}=5.4030\,\text{fm, }r_{t}=1.7494\,\text{fm, }v_{2t}=0.163\,\text{fm}^{3}%
\text{;} \\ 
a_{s}=-23.719\,\text{fm, }r_{s}=2.626\,\text{fm, }v_{2s}=-0.005\,\text{fm}%
^{3}
\end{array}
\label{18}
\end{equation}
differ significantly from the parameter values 
\begin{equation}
\begin{array}{c}
a_{t}=5.420\,\text{fm, }r_{t}=1.753\,\text{fm, }v_{2t}=0.040\,\text{fm}^{3}%
\text{;} \\ 
a_{s}=-23.739\,\text{fm, }r_{s}=2.678\,\text{fm, }v_{2s}=-0.48\,\text{fm}^{3}%
\text{,}
\end{array}
\label{19}
\end{equation}
\vspace{1pt}which were obtained on the basis of the data from the
partial-wave analysis of the Nijmegen group. The triplet low-energy
parameters calculated here for the phase shifts of the Nijmegen group are
virtually coincident with the analogous parameters obtained previously in
\lbrack 35\rbrack . Unfortunately, the data presented by the Nijmegen group
do not contain the singlet low-energy parameters of neutron--proton
scattering. The value of the singlet shape parameter $v_{2s}$ for the
Nijmegen phase shifts was calculated in \lbrack 1\rbrack , and it is in
agreement with our value.

Using expressions (18) and (19) for the scattering lengths and relying on
formulas (2) and (3), we find for the cross section $\sigma _{0}$ and for
the coherent scattering length $f$ that 
\begin{equation}
\sigma _{0}=20.426\,\text{b}\,,\text{ }f=-3.755\,\text{fm}  \label{20}
\end{equation}
in the case of the GWU phase shifts and that 
\begin{equation}
\sigma _{0}=20.473\,\text{b}\,,\text{ }f=-3.7395\,\text{fm}  \label{21}
\end{equation}
in the case of the Nijmegen phase shifts.

\vspace{1pt}

The values in (21) are in good agreement with the weighted mean of the cross
sections obtained by Houk and Dilg, $\sigma _{0}=20.476(12)\,$b, and with
the coherent-scattering-length value of $f=-3.7409(11)\,$fm according to
Koester and Nistler \lbrack 22\rbrack . It should be emphasized, however,
that this agreement is not accidental; it is directly related to the fact
that, in the partial-wave analysis of the Nijmegen group, the cross-section
values obtained by Houk \lbrack 12\rbrack\ and Dilg \lbrack 13\rbrack\ and
the coherent-scattering-length value obtained by Koester and Nistler \lbrack
22\rbrack\ were used as input experimental parameters. It is precisely the
reason why all of the experimental low-energy parameters in (17), with the
exception of the singlet effective range, agree within the experimental
error with the corresponding parameters in (19), which were calculated on
the basis of the Nijmegen phase shifts.

The singlet-effective-range value of $r_{s}=2.678\,$fm, which was calculated
for the phase shifts obtained by the Nijmegen group, is much smaller than
the experimental value of $r_{s}=2.77(5)\,$fm, which was quoted by Dilg in
\lbrack 13\rbrack . In this connection, it should be noted that, in \lbrack
13\rbrack , the singlet effective range $r_{s}$ was determined from
experimental data on the total cross section for neutron--proton scattering
at energies below $5\,$MeV at the scattering-length values fixed at $%
a_{t}=5.423(4)\,$fm and $a_{s}=-23.749\,$fm and the triplet-effective-range
value fixed at $r_{t}=1.760(5)\,$fm, but, as was indicated above, this
method for determining the effective range is highly unreliable (see Table
3). A determination of the singlet effective range $r_{s}$ directly from the
singlet phase shift irrespective of the triplet parameters is more correct
and consistent, which reduces substantially the uncertainty in this quantity.

For the sake of comparison, the low-energy parameters for neutron--proton
scattering that correspond to the GWU (GWU PWA) and Nijmegen (Nijm PWA)
phase shifts are given in Table 4, along with the values of these parameters
for a number of the realistic potentials (Argonne V18 \lbrack 37\rbrack ,
CD-Bonn \lbrack 28, 38\rbrack , and Moscow \lbrack 39\rbrack\ potentials)
whose parameters were matched with the Nijmegen nucleon--nucleon database.
Also quoted there are the experimental values of the low-energy parameters.
Table 4 shows that the values of the low-energy parameters obtained for the
Nijmegen phase shifts are in perfect agreement with the corresponding
parameters for the potentials fitted to the Nijmegen nucleon--nucleon
database.

A significant distinction between the values of the triplet low-energy
parameters for the GWU and Nijmegen data was discussed in detail in our
previous article \lbrack 4\rbrack . Here, we only indicate that the
difference of the triplet scattering lengths by $0.3\,\%$ is in fact a more
important circumstance than the fourfold distinction between the values of
the triplet shape parameters. This is because many important features of the
neutron--proton system --- such as the asymptotic deuteron normalization
factor $A_{S}$ and the root-mean-square radius $r_{d}$ of the deuteron ---
are highly sensitive to variations in the triplet scattering length \lbrack
40\rbrack . We also note that, although the triplet effective ranges
obtained from experimental data of the two main groups are close to each
other, the values of the difference $\delta r_{t}$ of the effective range $%
r_{t}$ and the mixed effective radius $\rho \left( -\varepsilon
_{d},0\right) $ for the GWU \lbrack 33\rbrack\ and Nijmegen \lbrack
34\rbrack\ phase shifts differ significantly. For example, the correction $%
\delta r_{t}$ for the phase shifts of the GWU group is positive, taking the
value 
\begin{equation}
\delta r_{t}=0.0163\,\text{fm}\,.  \label{22}
\end{equation}
For the phase shifts of the Nijmegen group, this correction is negative and,
in absolute value, is an order of magnitude smaller than the correction in
(22): $\delta r_{t}=-0.001\,$fm. The value of the singlet effective range
for the phase shifts of the GWU group also differs from its counterpart for
the Nijmegen phase shifts (by about $2\,\%$), and the corresponding
difference of the singlet shape parameters is formidable, reaching two
orders of magnitude.

In contrast to the partial-wave analysis of the Nijmegen group, the
partial-wave analysis of the GWU group does not employ the values of the
cross section $\sigma _{0}$ and the coherent scattering length $f$ as input
parameters. The theoretical values of $\sigma _{0}=20.426\,$b and $%
f=-3.755\, $fm, which we obtained here for the cross section in question and
for the neutron--proton coherent scattering length from data of the
partial-wave analysis performed by the GWU group, are in perfect agreement
with the experimental cross-section value of $\sigma _{0}=20.436(23)\,$b
according to Houk \lbrack 12\rbrack , and the experimental
coherent-scattering-length value obtained by Houk and Wilson \lbrack 9,
10\rbrack , 
\begin{equation}
f=-3.756(9)\,\text{fm\thinspace },  \label{23}
\end{equation}
but they contradict the cross-section value of $\sigma _{0}=20.491(14)\,$b
according to Dilg \lbrack 13\rbrack\ and the coherent-scattering-length
value of $f=-3.7384(20)\,$fm, which was obtained recently by the
neutron-interferometry method in \lbrack 23\rbrack . Thus, we see that a
reliable experimental determination of the total cross section for
neutron--proton scattering at zero energy, $\sigma _{0}$, and of the
coherent scattering length, $f$, is now quite a pressing problem. Precise
values of these quantities would make it possible to determine unambiguously
the triplet and singlet scattering lengths and to solve the problem of
choosing a correct set of the low-energy parameters and phase shifts among
currently recommended experimental values.

\vspace{1pt}

\begin{center}
REFERENCES
\end{center}

\begin{enumerate}
\item  T. D. Cohen and J. M. Hansen, Phys. Rev. C \textbf{59}, 13\ (1999).

\item  T. D. Cohen and J. M. Hansen, Phys. Rev. C \textbf{59}, 3047\ (1999).

\item  M. W. Kermode, A. McKerrell, J. P. McTavish, and L. J. Allen, Z.
Phys. A \textbf{303, }167 (1981).

\item  V. A. Babenko and N. M. Petrov, Yad. Fiz. \textbf{68}, 244 (2005)
\lbrack Phys. At. Nucl. \textbf{68}, 219 (2005)\rbrack .

\item  A. G. Sitenko and V. K. Tartakovski\u{\i}, \textit{Lectures on the
Theory of the Nucleus} (Atomizdat, Moscow, 1972; Pergamon, Oxford, 1975).

\item  L. Hulth\'{e}n and M. Sugawara, in \textit{Handbuch der Physik}, Ed.
by S. Fl\"{u}gge (Springer-Verlag, New York, Berlin, 1957), p. 1.

\item  E. Melkonian, Phys. Rev. \textbf{76}, 1744\ (1949).

\item  A. T. Stewart and G. L. Squires, Phys. Rev. \textbf{90}, 1125\ (1953).

\item  T. L. Houk and R. Wilson, Rev. Mod. Phys. \textbf{39}, 546\ (1967).

\item  T. L. Houk and R. Wilson, Rev. Mod. Phys. \textbf{40}, 672\ (1968).

\item  J. M. Neill, J. L. Russell, and J. R. Brown, Nucl. Sci. Eng. \textbf{%
33}, 265\ (1968).

\item  T. L. Houk, Phys. Rev. C\textbf{\ 3}, 1886\ (1971).

\item  W. Dilg, Phys. Rev. C\textbf{\ 11}, 103\ (1975).

\item  G. L. Squires and A. T. Stewart, Proc. Roy. Soc. A\textbf{\ 230}, 19\
(1955).

\item  J. Callerame, D. J. Larson, S. J. Lipson, and R. Wilson, Phys. Rev. C%
\textbf{\ 12}, 1423\ (1975).

\item  C. G. Shull, E. O. Wollan, G. A. Morton, and W. L. Davidson, Phys.
Rev. \textbf{73}, 842\ (1948).

\item  D. J. Hughes, M. T. Burgy, and G. R. Ringo, Phys. Rev. \textbf{77},
291\ (1950).

\item  M. T. Burgy, G. R. Ringo, and D. J. Hughes, Phys. Rev. \textbf{84},
1160\ (1951).

\item  W. C. Dickinson, L. Passell, and O. Halpern, Phys. Rev. \textbf{126},
632\ (1962).

\item  L. Koester, Z. Phys. \textbf{198}, 187 (1967).

\item  L. Koester and W. Nistler, Phys. Rev. Lett. \textbf{27}, 956\ (1971).

\item  L. Koester and W. Nistler, Z. Phys. A\textbf{\ 272}, 189 (1975).

\item  K. Schoen, D. L. Jacobson, M. Arif et al., Phys. Rev. C\textbf{\ 67},
044005\ (2003).

\item  O. Dumbrajs, R. Koch, H. Pilkuhn et al., Nucl. Phys. B\textbf{\ 216},
277\ (1983).

\item  H. P. Noyes, Phys. Rev. \textbf{130}, 2025\ (1963).

\item  H. P. Noyes, Ann. Rev. Nucl. Sci. \textbf{22}, 465 (1972).

\item  E. L. Lomon and R. Wilson, Phys. Rev. C\textbf{\ 9}, 1329\ (1974).

\item  R. Machleidt, Phys. Rev. C\textbf{\ 63}, 024001\ (2001).

\item  V. F. Kharchenko, N. M. Petrov, and S. A. Storozhenko, Nucl. Phys. A%
\textbf{\ 106, }464 (1968).

\item  V. F. Kharchenko, Fiz. \'{E}lem. Chastits At. Yadra \textbf{10}, 884\
(1979) \lbrack Sov. J. Part. Nucl. \textbf{10}, 349\ (1979)\rbrack .

\item  V. F. Kharchenko and S. A. Storozhenko, Nucl. Phys. A\textbf{\ 137, }%
437 (1969).

\item  N. M. Petrov and I. V. Simenog, Yad. Fiz. \textbf{28}, 381 (1978)
\lbrack Sov. J. Nucl. Phys. \textbf{28}, 193 (1978)\rbrack .

\item  R. A. Arndt, W. J. Briscoe, I. I. Strakovsky, and R. L. Workman, 
\textit{Partial-Wave Analysis Facility\ SAID}, The George Washington
University \lbrack http://gwdac.phys.gwu.edu\rbrack ; R. A. Arndt, I. I.
Strakovsky, and R. L. Workman, Phys. Rev. C \textbf{62}, 034005 (2000).

\item  Nijmegen NN-Online program \lbrack http://nn-online.org\rbrack ; V.
G. J. Stoks, R. A. M. Klomp, M. C. M. Rentmeester, and J. J. de Swart, Phys.
Rev. C\textbf{\ 48}, 792\ (1993).

\item  J. J. de Swart, C. P. F. Terheggen, and V. G. J. Stoks, \textit{%
Invited talk at the 3rd International Symposium ''Dubna Deuteron 95'',
Dubna, Russia, 1995,} nucl-th/9509032.

\item  V. G. J. Stoks, R. A. M. Klomp, C. P. F. Terheggen, and J. J. de
Swart, Phys. Rev. C\textbf{\ 49}, 2950\ (1994).

\item  R. B. Wiringa, V. G. J. Stoks, and R. Schiavilla, Phys. Rev. C\textbf{%
\ 51}, 38\ (1995).

\item  R. Machleidt, F. Sammarruca and Y. Song, Phys. Rev. C\textbf{\ 53},
1483\ (1996).

\item  V. I. Kukulin, V. N. Pomerantsev, and A. Faessler, nucl-th/9903056.

\item  V. A. Babenko and N. M. Petrov, Yad. Fiz. \textbf{66}, 1359 (2003)
\lbrack Phys. At. Nucl. \textbf{66}, 1319 (2003)\rbrack .
\end{enumerate}

\newpage

\noindent \textbf{Table 1.} Total cross section for neutron scattering on a
proton at zero energy

\begin{center}
\begin{tabular}{|c|c|c|}
\hline
$\,$No.$\,$ & References & $\sigma _{0}\,$, b \\ \hline
1 & \multicolumn{1}{|l|}{$\,$ Melkonian \lbrack 7\rbrack\ (1949)} & 
\multicolumn{1}{|l|}{$\,$ $20.36(10)$ $\,$} \\ 
2 & \multicolumn{1}{|l|}{$\,$ Stewart and Squires \lbrack 8\rbrack\ (1953) $%
\,$} & \multicolumn{1}{|l|}{$\,$ $20.41(14)$ $\,$} \\ 
3 & \multicolumn{1}{|l|}{$\,$ Houk and Wilson \lbrack 9\rbrack\ (1967)} & 
\multicolumn{1}{|l|}{$\,$ $20.37(2)$ $\,$} \\ 
4 & \multicolumn{1}{|l|}{$\,$ Houk and Wilson \lbrack 10\rbrack\ (1968)} & 
\multicolumn{1}{|l|}{$\,$ $20.442(23)$ $\,$} \\ 
5 & \multicolumn{1}{|l|}{$\,$ Neill et al. \lbrack 11\rbrack\ (1968)} & 
\multicolumn{1}{|l|}{$\,$ $20.366(76)$ $\,$} \\ 
6 & \multicolumn{1}{|l|}{$\,$ Houk \lbrack 12\rbrack\ (1971)} & 
\multicolumn{1}{|l|}{$\,$ $20.436(23)$ $\,$} \\ 
7 & \multicolumn{1}{|l|}{$\,$ Dilg \lbrack 13\rbrack\ (1975)} & 
\multicolumn{1}{|l|}{$\,$ $20.491(14)$ $\,$} \\ \hline
\end{tabular}
\end{center}

${}$

\newpage

\noindent \textbf{Table 2.} Amplitude for coherent neutron--proton scattering

\begin{center}
\begin{tabular}{|c|c|c|}
\hline
$\,$No.$\,$ & References & $f$, fm \\ \hline
1 & \multicolumn{1}{|l|}{$\,$ Shull et al. \lbrack 16\rbrack\ (1948)} & 
\multicolumn{1}{|l|}{$\,$ $-3.900(100)$ $\,$} \\ 
2 & \multicolumn{1}{|l|}{$\,$ Hughes et al. \lbrack 17\rbrack\ (1950)} & 
\multicolumn{1}{|l|}{$\,$ $-3.75(3)$} \\ 
3 & \multicolumn{1}{|l|}{$\,$ Burgy et al. \lbrack 18\rbrack\ (1951)} & 
\multicolumn{1}{|l|}{$\,$ $-3.78(2)$} \\ 
4 & \multicolumn{1}{|l|}{$\,$ Stewart and Squires \lbrack 8\rbrack\ (1955)}
& \multicolumn{1}{|l|}{$\,$ $-3.80(5)$} \\ 
5 & \multicolumn{1}{|l|}{$\,$ Dickinson et al. \lbrack 19\rbrack\ (1962)} & 
\multicolumn{1}{|l|}{$\,$ $-3.740(20)$ $\,$} \\ 
6 & \multicolumn{1}{|l|}{$\,$ Koester \lbrack 20\rbrack\ (1967)} & 
\multicolumn{1}{|l|}{$\,$ $-3.719(2)$} \\ 
7 & \multicolumn{1}{|l|}{$\,$ Houk and Wilson \lbrack 9, 10\rbrack\ (1967,
1968)} & \multicolumn{1}{|l|}{$\,$ $-3.756(9)$} \\ 
8 & \multicolumn{1}{|l|}{$\,$ Koester and Nistler \lbrack 21\rbrack\ (1971)}
& \multicolumn{1}{|l|}{$\,$ $-3.740(3)$} \\ 
9 & \multicolumn{1}{|l|}{$\,$ Koester and Nistler \lbrack 22\rbrack\ (1975)}
& \multicolumn{1}{|l|}{$\,$ $-3.7409(11)$ $\,$} \\ 
10 & \multicolumn{1}{|l|}{$\,$ Callerame et al. \lbrack 15\rbrack\ (1975)} & 
\multicolumn{1}{|l|}{$\,$ $-3.733(4)$} \\ 
11 & \multicolumn{1}{|l|}{$\,$ Schoen et al. \lbrack 23\rbrack\ (2003)} & 
\multicolumn{1}{|l|}{$\,$ $-3.7384(20)$ $\,$} \\ \hline
\end{tabular}
\end{center}

\newpage

\noindent \textbf{Table 3.} Low-energy parameters of neutron--proton
scattering from various studies

\vspace{1pt}\vspace{1pt}

\begin{center}
\begin{tabular}{|c|c|c|c|c|c|}
\hline
$\,$No.$\,$ & References$\,$ & $a_{t}\,,\,$fm & $a_{s}\,,\,\,$fm & $%
r_{t}\,,\,\,$fm & $r_{s}\,,\,$fm \\ \hline
1 & \multicolumn{1}{|l|}{$\,$ Burgy et al. \lbrack 18\rbrack\ (1951)} & $\,$ 
$5.377(21)$ $\,$ & $\,$ $-23.690(55)$ $\,$ & $\,$ $1.704(28)$ $\,$ & $-$ \\ 
\hline
2 & \multicolumn{1}{|l|}{$\,$ Noyes \lbrack 25\rbrack\ (1963)} & $5.396(11)$
& $-23.678(28)$ & $1.727(14)$ & $\,$ $2.51(11)$ $\,$ \\ \hline
& \multicolumn{1}{|l|}{$\,$} & $5.392(6)$ & $-23.689(13)$ & $1.724(7)$ & $%
2.42(9)$ \\ 
3 & \multicolumn{1}{|l|}{$\,$ Houk and Wilson \lbrack 9\rbrack\ (1967) $\,$}
& $\,$ $5.399(11)$ $\,$ & $-23.680(28)$ & $1.732(12)$ & $2.48(11)$ \\ 
& \multicolumn{1}{|l|}{$\,$} & $5.411(4)$ & $-23.671(12)$ & $1.747(4)$ & $%
2.59(8)$ \\ \hline
4 & \multicolumn{1}{|l|}{$\,$ Houk and Wilson \lbrack 10\rbrack\ (1968)} & $%
5.405(6)$ & $-23.728(13)$ & $1.738(7)$ & $2.56(10)$ \\ \hline
5 & \multicolumn{1}{|l|}{$\,$ Koester and Nistler \lbrack 21\rbrack\ (1971) $%
\,$} & $5.414(5)$ & $-23.719(13)$ & $-$ & $-$ \\ \hline
6 & \multicolumn{1}{|l|}{$\,$ Noyes \lbrack 26\rbrack\ (1972)} & $5.413(5)$
& $-23.719(13)$ & $1.735$ & $2.66$ \\ 
& \multicolumn{1}{|l|}{$\,$} & $5.423(5)$ & $-23.712(13)$ & $1.748(6)$ & $%
2.75(10)$ \\ \hline
7 & \multicolumn{1}{|l|}{$\,$ Lomon and Wilson \lbrack 27\rbrack\ (1974) $\,$%
} & $5.414(5)$ & $-23.719(13)$ & $1.750(5)$ & $2.76(5)$ \\ \hline
& \multicolumn{1}{|l|}{$\,$} &  &  &  & $2.77(5)$ \\ 
8 & \multicolumn{1}{|l|}{$\,$ Dilg \lbrack 13\rbrack\ (1975)} & $5.423(4)$ & 
$-23.749(9)$ & $1.760(5)$ & $2.81(5)$ \\ 
& \multicolumn{1}{|l|}{$\,$} &  &  &  & $2.78(5)$ \\ \hline
9 & \multicolumn{1}{|l|}{$\,$ Koester and Nistler \lbrack 22\rbrack\ (1975) $%
\,$} & $5.424(3)$ & $-23.749(8)$ & $1.760(5)$ & $2.81(5)$ \\ \hline
10 & \multicolumn{1}{|l|}{$\,$ Dumbrajs et al. \lbrack 24\rbrack\ (1983) $\,$%
} & $5.424(4)$ & $-23.748(10)$ & $1.759(5)$ & $2.75(5)$ \\ \hline
11 & \multicolumn{1}{|l|}{$\,$ Machleidt \lbrack 28\rbrack\ (2001)} & $%
5.419(7)$ & $-23.740(20)$ & $1.753(8)$ & $2.77(5)$ \\ \hline
\end{tabular}
\\[0pt]

\vspace{1pt}\newpage
\end{center}

%\vspace{1pt}
\noindent \textbf{Table 4.} Low-energy parameters of neutron--proton
scattering that were obtained on the basis of the present-day data of the
partial-wave analysis and modern realistic models of nucleon--nucleon
interaction

\vspace{1pt}\vspace{1pt}

\begin{center}
\begin{tabular}{|c|c|c|c|c|c|c|c|}
\hline
$\,$No.$\,$ & Model$\,$ & $a_{t}\,,\,$fm & $a_{s}\,,\,\,$fm & $r_{t}\,,\,\,$%
fm & $r_{s}\,,\,$fm & $\sigma _{0}\,$, b & $f$, fm \\ \hline
1 & \multicolumn{1}{|l|}{$\,$ GWU PWA $\,$} & $5.4030$ & $-23.719$ & $1.7494$
& $2.626$ & $20.426$ & $-3.755$ \\ 
2 & \multicolumn{1}{|l|}{$\,$ Nijm PWA $\,$} & $5.420$ & $-23.739$ & $1.753$
& $2.678$ & $20.473$ & $-3.7395$ \\ 
3 & \multicolumn{1}{|l|}{$\,$ Argonne V18 $\,$} & $5.419$ & $-23.732$ & $%
1.753$ & $2.697$ & $20.461$ & $-3.7375$ \\ 
4 & \multicolumn{1}{|l|}{$\,$ CD Bonn $\,$} & $5.4199$ & $-23.738$ & $1.751$
& $2.671$ & $20.471$ & $-3.7392$ \\ 
5 & \multicolumn{1}{|l|}{$\,$ Moscow $\,$} & $5.422$ & $-23.740$ & $1.754$ & 
$2.66$ & $20.476$ & $-3.7370$ \\ 
6 & \multicolumn{1}{|l|}{$\,$ Expt. \lbrack 10, 12\rbrack} & $\,5.405(6)\,$
& $\,-23.728(13)\,$ & $\,1.738(7)\,$ & $\,2.56(10)\,$ & $\,20.436(23)\,$ & $%
-3.756(9)$ \\ 
7 & \multicolumn{1}{|l|}{$\,$ Expt. \lbrack 24\rbrack} & $5.424(4)$ & $%
-23.748(10)$ & $1.759(5)$ & $2.75(5)$ & $20.491(14)$ & $-3.738(1)$ \\ 
8 & \multicolumn{1}{|l|}{$\,$ Expt. \lbrack 28\rbrack} & $5.419(7)$ & $%
-23.740(20)$ & $1.753(8)$ & $2.77(5)$ & $20.476(12)$ & $\,-3.7409(11)\,$ \\ 
\hline
\end{tabular}
\\[0pt]

\vspace{1pt}
\end{center}

\end{document}